\documentclass[a4paper,11pt]{article}
\usepackage{pos}

\renewcommand{\d}{\ensuremath{\mathrm d} }
\newcommand{\cD}{\ensuremath{\mathcal D} }
\newcommand{\cO}{\ensuremath{\mathcal O} }
\newcommand{\de}{\ensuremath{\delta} }
\newcommand{\vev}[1]{\ensuremath{\left\langle #1 \right\rangle} }
\newcommand{\eq}[1]{Eq.~\ref{#1}}
\newcommand{\fig}[1]{Fig.~\ref{#1}}
\newcommand{\secref}[1]{Section~\ref{#1}}
\newcommand{\refcite}[1]{Ref.~\cite{#1}}

\title{Density of states for gravitational waves}

\author*{Felix Springer}
\author{David Schaich}

\affiliation{Department of Mathematical Sciences, University of Liverpool, \\ Liverpool L69 7ZL, United Kingdom}

\emailAdd{felix.springer@liverpool.ac.uk}
\emailAdd{david.schaich@liverpool.ac.uk}

\FullConference{38th International Symposium on Lattice Field Theory \\ 26--30 July 2021 \\ Zoom/Gather @ Massachusetts Institute of Technology}

\abstract{We present ongoing investigations of the first-order confinement transition of a composite dark matter model, to predict the resulting spectrum of gravitational waves.
To avoid long autocorrelations at the first-order transition, we employ the Logarithmic Linear Relaxation (LLR) density of states algorithm.
After testing our calculations by reproducing existing results for compact U(1) lattice gauge theory, we focus on the pure-gauge SU(4) theory related to the Stealth Dark Matter model.}

\begin{document}
\maketitle

\section{Introduction} 
Standard Markov-chain Monte Carlo techniques have proven to be an invaluable tool in modern theoretical physics. However there are situations in which these standard techniques fall short, and alternatives such as density of states approaches may perform better. These include lattice field theory studies of first-order phase transitions, where uncontrollable autocorrelations can result from the difficulty of tunnelling between the two coexisting phases on large lattice volumes.

In this work we are interested in the gravitational-wave signatures that arise from such first-order phase transitions in composite dark matter models. In particular, we consider the Stealth Dark Matter model being investigated by the Lattice Strong Dynamics Collaboration~\cite{Appelquist:2015yfa, Appelquist:2015zfa, LatticeStrongDynamics:2020jwi}. This is an SU(4) gauge theory coupled to four fermions that transform in the fundamental representation of the gauge group. While these fundamental fermions would be electrically charged, the resulting `dark baryon' formed by them would be an electroweak-singlet scalar particle.

In addition to guaranteeing the stability of a massive dark matter candidate through an analogue of baryon number conservation, the symmetries of Stealth Dark Matter strongly suppress its scattering cross section in direct-detection experiments~\cite{Appelquist:2015zfa, Appelquist:2015yfa}, especially for heavy dark matter masses $M_{\text{DM}} \gtrsim 1$~TeV.
Collider searches for Stealth Dark Matter also become challenging for such heavy masses~\cite{Kribs:2018ilo, Butterworth:2021jto}, which motivates ongoing work~\cite{LatticeStrongDynamics:2020jwi, Huang:2020mso, Kang:2021epo} investigating the possibility that such models could be constrained or discovered by future gravitational-wave observatories.

At the high temperatures of the early universe, the `dark gluons' and `dark fermions' would exist in a deconfined plasma, and as the universe cools the model would transition to a confined phase. For sufficiently heavy fermions (relative to the confinement scale), this phase transition would be first order and would therefore generate a stochastic background of gravitational waves, which space-based observatories such as LISA will search for~\cite{Caprini:2019egz}.

In this proceedings we report on ongoing work applying the Logarithmic Linear Relaxation (LLR) density of states algorithm~\cite{Langfeld:2012ah, Langfeld:2015fua} to analyze pure-gauge SU(4) Yang--Mills theory, which can be considered the `quenched' limit of Stealth Dark Matter corresponding to infinitely heavy fermions.
Pure-gauge theory is convenient for the LLR algorithm, which is challenging to apply to systems with dynamical fermions~\cite{Korner:2020vjw}.
The SU($N$) Yang--Mills confinement transition is known to be first order for $N \geq 3$, with significantly stronger transitions for $N \geq 4$~\cite{Lucini:2012gg}.
This makes the SU(4) theory a promising first target for application of the LLR algorithm, which future work could extend either to the SU(3) case relevant for QCD or to $N \geq 5$ to explore the large-$N$ limit.

We begin in the next section by summarizing the LLR algorithm.
In \secref{U(1)} we discuss how we tested our LLR code by reproducing some results for compact U(1) lattice gauge theory from \refcite{Langfeld:2015fua}.
Then we present initial results from our ongoing LLR analyses of the SU(4) Yang--Mills confinement transition in \secref{SU(4)}, wrapping up in \secref{summary} with a brief overview of the next steps we have planned.

\section{Algorithm to determine the density of states} 
In SU($N$) lattice Yang--Mills theories, observables are defined by
\begin{align}
  \label{eq:obs}
  \vev{\cO} & = \frac{1}{Z} \int \cD\phi \, \cO(\phi) \, e^{-S[\phi]} &
  Z & = \int \cD \phi \, e^{-S[\phi]},
\end{align}
with $\phi$ the set of field variables and $S[\phi]$ the lattice action. In standard techniques the idea is to approximate this extremely high-dimensional path integral by sampling a modest number of field configurations according to a probability $\propto e^{-S[\phi]}$.

Here our aim is to calculate the density of states
\begin{equation}
  \rho(E) = \int \cD \phi \, \delta(S[\phi] - E)
\end{equation}
and use it to rewrite \eq{eq:obs} as a one-dimensional integral over the energy,
\begin{align}
  \vev{\cO(\beta)} & = \frac{1}{Z(\beta)} \int \d E \, \cO(E) \, \rho(E) \, e^{\beta E} &
  Z(\beta) & = \int \d E \, \rho(E) \, e^{\beta E}.
\end{align}
The challenge now lies in determining the density of states $\rho(E)$.
The algorithm we are using to do this is called Logarithmic Linear Relaxation (LLR)~\cite{Langfeld:2012ah, Langfeld:2015fua}.
The procedure is the following: First we define the reweighted expectation value
\begin{align}
  \vev{\vev{E - E_i}}_{\de}(a) & = \frac{1}{N}\int \cD \phi \, (E-E_i) \, \theta_{E_i,\de} \, e^{-aS[\phi]} = \frac{1}{N} \int_{E_i-\frac{\de}{2}}^{E_i+\frac{\de}{2}} \d E \, (E-E_i) \, \rho(E) \, e^{-aE}, \label{Heavyside} \\
  N & = \int \cD \phi \, \theta_{E_i,\de} \, e^{-aS[\phi]} = \int_{E_i-\frac{\de}{2}}^{E_i+\frac{\de}{2}} \d E \, \rho(E) \, e^{-a E},
\end{align}
where $E_i$ are a set of fixed energy values, $\theta_{E_i,\de}$ is the modified Heaviside function ($1$ in the interval $E_i \pm \frac{\delta}{2}$ and $0$ everywhere else), and for now `$a$' is just a free parameter not to be confused with the lattice spacing.

Next we set the reweighted expectation value $\vev{\vev{E - E_i}}_{\de}(a)$ to zero and rewrite the integral using the trapezium rule:
\begin{align}
  \vev{\vev{E - E_i}}_{\de}(a) & = \frac{1}{N} \int_{E_i-\frac{\de}{2}}^{E_i+\frac{\de}{2}} \d E \, (E-E_i) \, \rho(E) \, e^{-aE} \\
  & =\frac{1}{N} \frac{\de}{2}\left( (\frac{\de}{2})e^{-a(E_i+\frac{\delta}{2})}\rho(E_i+\frac{\de}{2})+(-\frac{\de}{2})e^{-a(E_i-\frac{\de}{2})}\rho(E_i-\frac{\de}{2})\right) + \cO(\de^3) = 0. \nonumber
\end{align}
Now we rewrite the exponential $e^{\pm a \frac{\de}{2}}$ and the density $\rho(E_i\pm \de/2)$ as a Taylor series expansion, and neglect $\cO(\de^2)$ terms by considering $\de \rightarrow 0$:
\begin{align}
  0 & = \left(\rho(E_i) + \frac{\de}{2} \frac{\d \rho(E)}{\mathrm{d}E}\Bigr|_{E=E_i}\right)\left(1-a\frac{\de}{2}\right) - \left(\rho(E_i)-\frac{\de}{2}\frac{\d \rho(E)}{\d E}\Bigr|_{E=E_i}\right)\left(1+a\frac{\de}{2}\right) \nonumber \\
  & = \left(-\rho(E_i)a + \frac{\d \rho(E)}{\d E}\Bigr|_{E = E_i} - \rho(E_i)a + \frac{\d \rho(E)}{\d E}\Bigr|_{E=E_i}\right)\frac{\de}{2} \\
  \implies a & = \frac{1}{\rho(E_i)}\frac{\d \rho(E)}{\d E}\Bigr|_{E=E_i} = \frac{\d \ln(\rho(E))}{\d E}\Bigr|_{E=E_i}. \label{eq:a}
\end{align}
We now have an interpretation of $a$ as the derivative of the logarithm of the density of states, and a way to obtain $\rho(E)$ within each interval $E_i \pm \frac{\de}{2}$ by numerically integrating values of $a(E_i)$.

A key feature of this algorithm is that the derivative in \eq{eq:a} is determined without taking differences. As \refcite{Langfeld:2016kty} reviews, this makes it possible to exponentially suppress errors, enabling the calculation of the density of states $\rho(E)$ over many orders of magnitude with nearly constant relative uncertainties.

To compute $a$ for a given $E_i$, we solve the equation $\vev{\vev{E - E_i}}_{\de}(a) = 0$ using the Robbins--Monro algorithm~\cite{Langfeld:2015fua}:
\begin{equation}
  a^{(n+1)}=a^{(n)}+ \frac{12}{\de^2 (n+1)}\vev{\vev{E - E_i}}_{\de}(a^{(n)}),
\end{equation}
which has a fixed point at the correct value of $a=a^{(n+1)}=a^{(n)}$.
For the evaluation of the reweighted expectation value $\vev{\vev{E - E_i}}_{\de}(a^{(n)})$ standard importance-sampling Monte Carlo techniques are used, but with the weight $e^{-a^{(n)}S}$ rather than the usual $e^{-S}$.

The use of the modified Heaviside function $\theta_{E_i,\de}$ in \eq{Heavyside} implies that we reject all Monte Carlo updates that would produce a configuration with an energy outside of $E_i \pm \frac{\delta}{2}$. This can lead to low acceptance rates for small energy intervals $\delta$. An alternative is to use a smooth Gaussian window function instead of this hard cut-off~\cite{Langfeld:2016kty, Korner:2020vjw}:
\begin{equation}
  \label{eq:gauss}
  \vev{\vev{E - E_i}}_{\de}(a) = \frac{1}{N} \int \d E \, (E-E_i) \, \exp\left[-\frac{(E - E_i)}{2\de^2}\right] \, \rho(E) \, e^{-aE}.
\end{equation}
In our work we experiment with both of these options.

\section{Checking the compact U(1) transition} \label{U(1)} 
To test our implementation of the LLR algorithm, we first reproduce results from a previous study of 4d pure-gauge U(1) theory, Ref.~\cite{Langfeld:2015fua}. The action of this model is
\begin{equation}
  S = -\beta \sum_{x,\mu < \nu} \cos(\theta_{\mu \nu}(x)),
\end{equation}
with $\theta_{\mu \nu}(x) = \theta_{\mu}(x) + \theta_{\nu}(x + \hat{\mu}) - \theta_{\mu}(x+\hat{\nu}) - \theta_{\nu}(x)$.
Here $\beta = \frac{1}{g_0^2}$ with $g_0^2$ the bare gauge coupling, the sum runs over all lattice sites, and $\theta_{\mu}(x) \in [-\pi; \pi]$ is the compact variable of the link attached to lattice site $x$ in direction $\hat{\mu}$.

\begin{figure}
  \includegraphics[width=7.1cm]{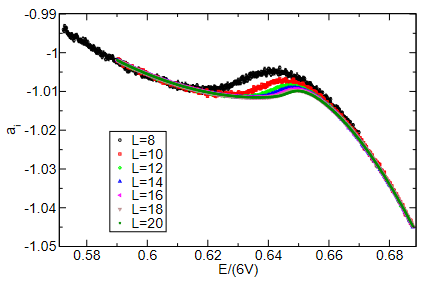}\hfill
  \includegraphics[width=7.1cm]{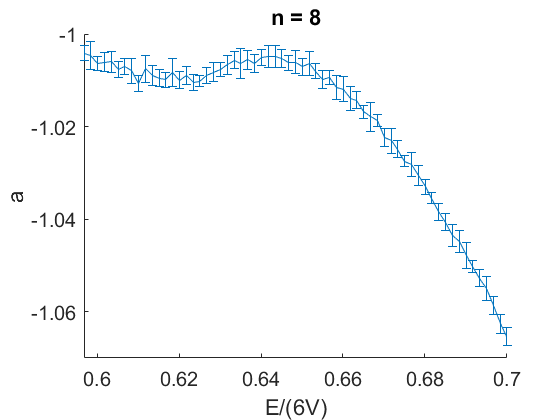}
  \caption{Comparing results for $a$ obtained for compact U(1) pure-gauge theory. The left plot is from \refcite{Langfeld:2015fua}, for lattice volumes $V$ between $8^4$ and $20^4$ and energy intervals of size $\de = 0.0011457/V$. The right plot displays consistent values calculated by us for $V=8^4$, $\de = 0.001/V$ and $N_{\text{Jackknife}}=10$.}
  \label{fig:comparison}
\end{figure}

We ran our LLR calculations using a lattice volume $V=8^4$, with an energy interval size of $\de = 0.001/V$ and $N_{\text{Jackknife}}=10$ independent runs of the Robbins--Monro algorithm for each energy interval.
For this U(1) case, we use the hard energy cut-off of \eq{Heavyside}, meaning that we reject every Monte Carlo update that would produce a lattice configuration with an energy outside the interval $E_i \pm \frac{\de}{2}$. As illustrated by \fig{fig:comparison}, our results for the values of $a$ are consistent with the results from \refcite{Langfeld:2015fua}, given the different \de employed in each case.
This provides confidence that our underlying implementation of the LLR algorithm is working correctly.

\begin{figure}
  \centering
  \includegraphics[width=6cm]{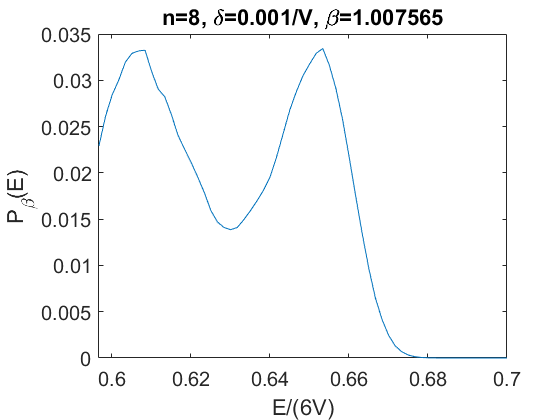}
  \caption{Probability density $P_{\beta}(E) = \rho(E) e^{\beta E}$ (omitting uncertainties) for compact U(1) lattice gauge theory, using lattice volume $V=8^4$, energy interval size $\delta = 0.001/V$, and $\beta = 1.007565$ close to the critical value.}
  \label{fig:rho_U1}
\end{figure}

Using these results for $a$, and choosing $\beta=0.9965$ close to the critical value for compact U(1) lattice gauge theory, we obtain the probability density $P_{\beta}(E) = \rho(E) e^{\beta E}$ shown in \fig{fig:rho_U1}.
For the reconstruction of the density of states $\rho$ from the values of $a$ in this case, we used a simple trapezium rule numerical integration.
The distribution in \fig{fig:rho_U1} has a clear double-peak structure, signalling the coexistence of the two phases at the first-order bulk phase transition.
From this plot we can directly read off the latent heat $\Delta E / (6V) \approx 0.047$ as the separation between the two peaks.

\section{SU(4) Yang--Mills confinement transition} \label{SU(4)} 
We now proceed to apply the LLR algorithm to SU(4) Yang--Mills theory. The action is
\begin{equation}
  \label{eq:wilson}
  S = -\beta \sum_{x,\mu<\nu} \mathrm{Re}\mathrm{Tr}\left(U_{\mu\nu}(x)\right),
\end{equation}
with the plaquette $U_{\mu\nu}(x) = U_{\mu}(x) U_{\nu}(x+\hat{\mu}) U_{\mu}^{\dagger}(x+\hat{\nu}) U_{\nu}^{\dagger}(x)$.
Here $\beta = \frac{8}{g_0^2}$, with $g_0^2$ the bare gauge coupling, the sum runs over all lattice sites and $U_{\mu}(x)$ is the SU(4)-valued link variable of the link attached to lattice site $x$ in direction $\hat{\mu}$.

We implemented the LLR algorithm in a fork of Stefano Piemonte's \texttt{LeonardYM} software.\footnote{\texttt{\href{https://github.com/spiemonte/LeonardYM}{github.com/spiemonte/LeonardYM}}; \quad \texttt{\href{https://github.com/FelixSpr/LeonardYM}{github.com/FelixSpr/LeonardYM}} \\ See also \texttt{\href{https://github.com/milc-qcd/milc_qcd/issues/44}{github.com/milc-qcd/milc\_qcd/issues/44}}}
In addition to supporting SU($N$) Yang--Mills theories with arbitrary $N$, \texttt{LeonardYM} also offers efficient MPI+OpenMP data parallelism.
However, because we use sweeps of overrelaxation updates in the full SU(4) group~\cite{Creutz:1987xi} to calculate the reweighted expectation value $\vev{\vev{E - E_i}}_{\de}(a^{(n)})$ for every Robbins--Monro iteration, we are currently not parallelizing the lattice volume.
This is due to the appearance of the global energy $E$ in these updates, for both the hard energy cut-off of \eq{Heavyside} and the Gaussian window of \eq{eq:gauss}.
At present we obtain sufficient parallelism from running $N_{\text{Jackknife}} = 4$ independent calculations for each energy interval.
Switching from overrelaxation to hybrid Monte Carlo updates in the future would allow data parallelism to be added, which may be necessary to reach lattice volumes too large for a single core to handle.

\begin{figure}
  \centering
  \includegraphics[width=12cm]{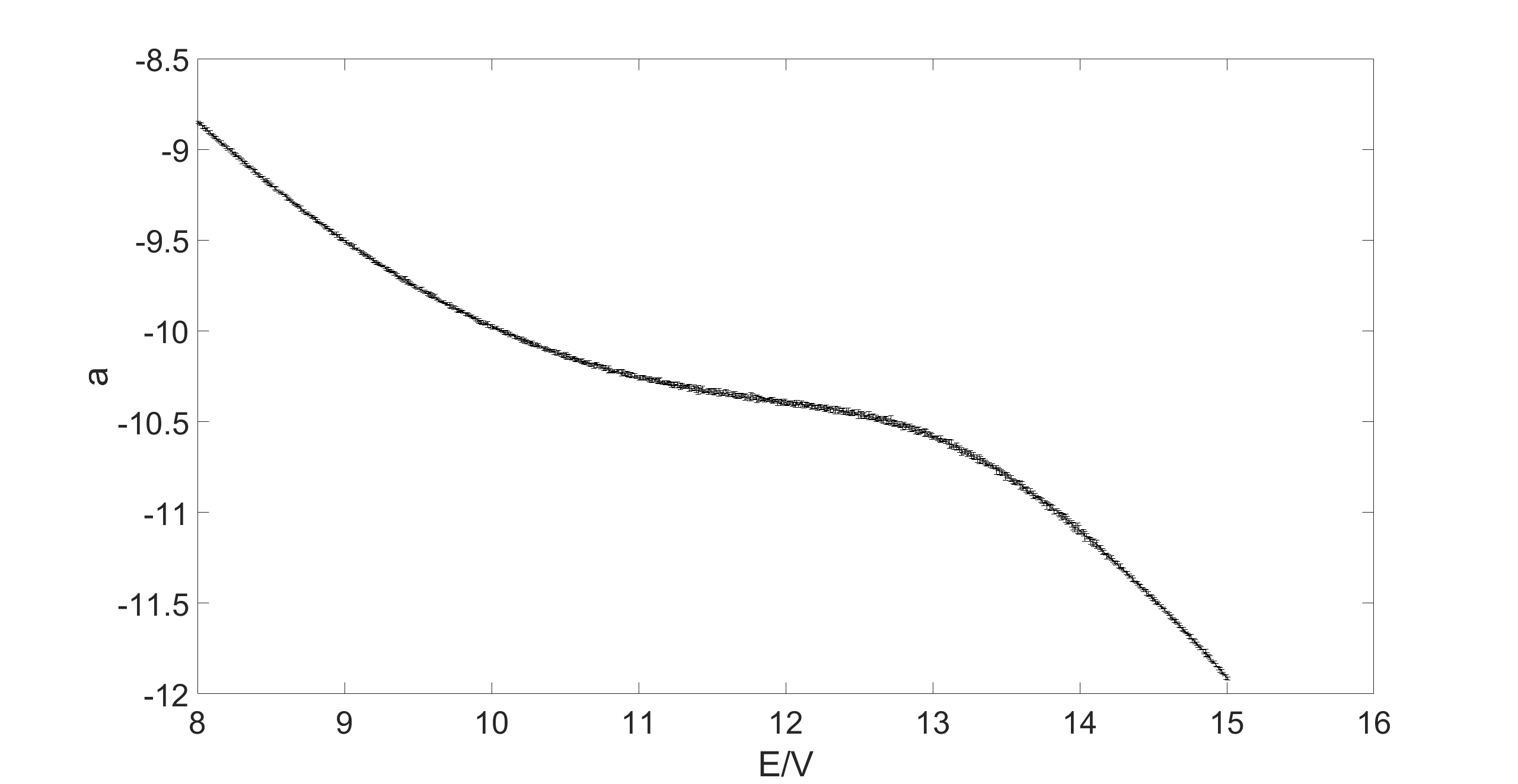}
  \caption{SU(4) results for $a$ vs.\ the energy density $E/V$, for a lattice of size $V=12^3 \times 6$ and an interval size of $\delta = 0.01/V$.  The error bars estimated by performing $N_{\text{Jackknife}} = 4$ independent runs per interval are very small.}
  \label{fig:avse}
\end{figure}

The largest volume we have considered so far is $V=12^3 \times 6$, with an energy interval size of $\delta = 0.01/V$ and $N_{\text{Jackknife}} = 4$ samples per energy interval.
Figure~\ref{fig:avse} shows our results for $a$ from roughly $700$ energy intervals spanning the range $8 \lesssim E/V \lesssim 15$.
In these calculations we use the Gaussian window function of \eq{eq:gauss}, as opposed to the hard cut-off we implemented for the U(1) case.
The smoother window helps to keep acceptance rates reasonably high---around $70\%$, much larger than we obtained using a hard cut-off.\footnote{It may also be possible to improve acceptance rates when using a hard cut-off by more cleverly constructing the overrelaxation updates~\cite{Kiskis:2003rd, deForcrand:2005xr}.}

From the results for $a$ shown in \fig{fig:avse}, we compare two methods to reconstruct the density of states $\rho(E)$.
The first of these is the same simple trapezium rule numerical integration we employed in the U(1) case.
The second fits the results to simple polynomials, which Refs.~\cite{Francesconi:2019aet, Francesconi:2019nph} argue is a more robust way to quantify and control systematic uncertainties from the reconstruction.
Because $\rho(E) = \rho(-E)$, only odd powers of $E$ are included in the $L$-term fit function $a_L(E)$, which is then integrated and exponentiated to obtain the density of states:
\begin{align}
  a_L(E) & = \sum_{i=1}^L c_i E^{2i - 1} &
  \longrightarrow & &
  \rho_L(E) & = \exp\left[\sum_{i=1}^L \frac{c_i}{2i} E^{2i}\right],
\end{align}
normalized such that $\rho_L(0) = 1$.
We choose to include $L = 8$ terms in our fits by varying $L$ and monitoring the resulting $\chi^2 / \text{dof}$.

\begin{figure}
  \centering
  \includegraphics[width=15cm]{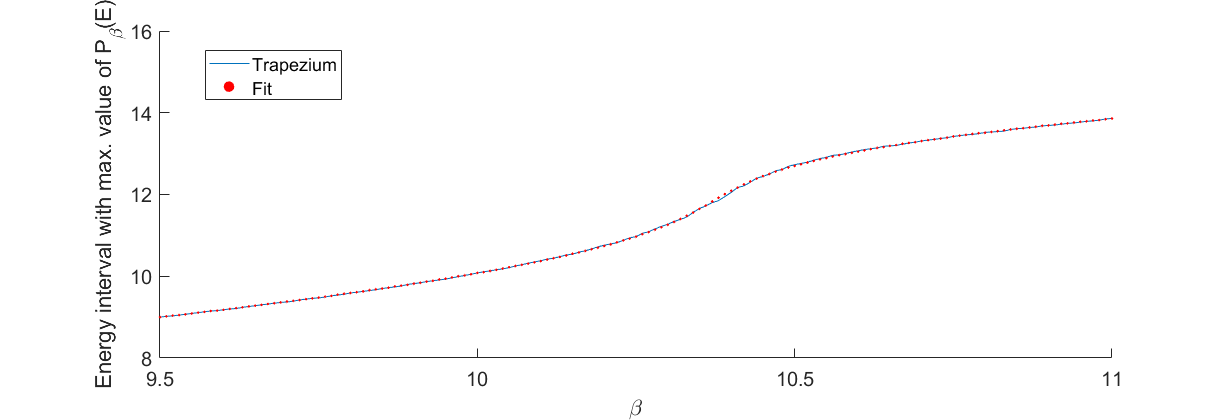}
  \caption{Energy interval containing the maximum value of $P_{\beta}(E) = \rho(E) e^{\beta E}$ vs.\ $\beta$, for $12^3 \times 6$ lattices, comparing two different methods of reconstructing the density of states from the results for $a$ in \fig{fig:avse}.  The rapid change around $\beta \approx 10.4$ corresponds to the SU(4) bulk transition.}
  \label{peakvsbeta}
\end{figure}

At present we see no significant difference between these two ways of reconstructing the density of states from our results for $a$.
This is illustrated by \fig{peakvsbeta}, in which we plot the energy interval containing the maximum value of the probability density $P_{\beta}(E) = \rho(E) e^{\beta E}$ across a broad range of $9.5 \leq \beta \leq 11$.
We consider this quantity because our SU(4) results do not yet produce any two-peaked probability distributions like that in \fig{fig:rho_U1} for the U(1) case.
We therefore identify pseudo-critical couplings by observing where the single peak in the distribution $P_{\beta}(E)$ moves most rapidly as $\beta$ changes.

In \fig{peakvsbeta} we are able to resolve only a single pseudo-critical region corresponding to the SU(4) bulk transition with $10.2 \lesssim \beta_{\text{bulk}} \lesssim 10.5$.
Based on results from \refcite{Wingate:2000bb}, which also uses the Wilson action \eq{eq:wilson}, we expect $N_t = 6$ to produce $\beta_c \approx 10.8$ for the physically relevant confinement transition, well separated from this bulk transition.
It is possible that the $12^3 \times 6$ lattices we have considered so far are simply too small to reveal the expected first-order confinement transition.
In addition, \refcite{Wingate:2000bb} also finds a latent heat $\Delta E / V \approx 0.004$ for $16^3 \times 6$ lattices, which may be too small to resolve using our current energy interval size $\de = 0.01 / V$.\footnote{This latent heat can be contrasted with $\Delta E / V \approx 0.28$ from \fig{fig:rho_U1} for the U(1) bulk transition.}
These considerations provide some clear next steps for our work, which we discuss in the next section.

\section{Outlook and next steps}\label{summary} 
In this proceedings we have presented our progress applying the LLR density of states algorithm to investigate first-order transitions in lattice gauge theories.
Standard Markov-chain importance-sampling techniques can suffer from long autocorrelations at such first-order transitions, which we aim to avoid by instead analyzing the density of states.
We have successfully tested our methods by reproducing results from \refcite{Langfeld:2015fua} for the first-order bulk transition of compact U(1) lattice gauge theory.
Motivated by the Stealth Dark Matter model and the stochastic spectrum of gravitational waves that would result from a first-order dark-sector confinement transition in the early Universe, our ongoing work focuses on SU(4) Yang--Mills theory.

So far our SU(4) computations with lattice volumes up to $12^3 \times 6$ suffice only to resolve the bulk transition, which for $N_t = 6$ is well separated from the first-order confinement transition of interest~\cite{Wingate:2000bb}.
We are currently improving our analyses by reducing the size of the energy intervals, to $\delta = 0.001 / V$.
To continue considering only hundreds (and not thousands) of independent energy intervals, this requires focusing our calculations on a smaller range of energies around $E / V \approx 13$.
In a similar vein, we will also analyze larger $N_s^3 \times N_t$ lattice volumes, both increasing $N_s$ with fixed $N_t$ to study the thermodynamic limit and increasing $N_t$ with fixed aspect ratio $N_s / N_t$ to extrapolate to the continuum limit.
Once we have resolved the first-order confinement transition, we will be able to predict observables such as the latent heat and interface tension that affect the resulting spectrum of gravitational waves.

Following the completion of this work, we will also be in a good position to generalize our studies beyond SU(4) Yang--Mills theory.
For example, it may be interesting to turn to the weakly first-order confinement transition of SU(3) Yang--Mills theory corresponding to quenched QCD.
We can also consider SU($N$) theories with $N \geq 5$ to investigate the behavior of this transition in the large-$N$ limit.
A more ambitious future target would be to explore the challenge of including fermion fields in LLR density of states calculations, which would allow us to consider the full Stealth Dark Matter model as in Ref.~\cite{LatticeStrongDynamics:2020jwi}.
In all of these cases, the presence of first-order transitions gives us reason to expect LLR density of states analyses to provide an advantage over standard importance sampling.

\vspace{20 pt} 
\noindent \textsc{Acknowledgments:}~We thank Kurt Langfeld, Paul Rakow, David Mason, James Roscoe and Johann Ostmeyer for helpful conversations about the LLR algorithm.
We also thank Georg Bergner and Stefano Piemonte for advice about \texttt{LeonardYM}.
Numerical calculations were carried out at the University of Liverpool.
DS was supported by UK Research and Innovation Future Leader Fellowship {MR/S015418/1} and STFC grant {ST/T000988/1}.

\bibliographystyle{JHEP}
\bibliography{lattice21}
\end{document}